# PRomop: A Decision-Ready Longitudinal Patient Health Record on the OMOP Common Data Model

**Adam Blum***[1] **Louis Ferger-Andrews**[2,3], **Steven Labkoff, MD**[2,4]

[1] HealthKey, Inc., South Jordan, UT, USA

[2] Division of Clinical Informatics, Beth Israel Deaconess Medical Center, Boston, MA 02115, USA

[3] Fontys University of Applied Sciences, Venlo, Netherlands

[4] Luminant Consulting, Stamford, CT 06903, USA

* Corresponding Author: Adam Blum, Chestnut House, Melville Castle, Lasswade, UK EH18 1AW. Email: adam@healthkey.ai

ORCID iDs: Adam Blum, https://orcid.org/0009-0009-4985-7615; Louis Ferger-Andrews, https://orcid.org/0009-0005-7909-0241; Steven Labkoff, https://orcid.org/0000-0002-2337-5185

## Abstract

**Objective:** Health systems and biopharma face a persistent gap between holding patient data and acting on it: records are fragmented, manually mapped, and structured for storage rather than decisions, so every application re-derives patient state. We present PRomop, an open-source longitudinal record that closes this gap.

**Materials and Methods:** PRomop builds on the OMOP Common Data Model (CDM 5.4) with oncology extensions and adds PatientRecord, a flattened projection collapsing each patient's longitudinal history into a single decision-ready 304-column row. State derivations—lines of therapy, disease status, normalized biomarkers—are computed once at projection time, so analytics, trial matching, and standard-of-care evaluation read one shared substrate.

**Results:** PRomop is deployed by two oncology organizations—the independently governed HealthTree Foundation (~14,000 patients) and CancerBot (~3,500), a HealthKey-owned deployment—matching against 19,500 recruiting trials across five cancer types. A 20-criterion eligibility search requiring 27–39 joins over raw OMOP reduces to zero against the projection. On a synthetic 1000-patient breast-cancer cohort, eligibility screening averaged 0.30 ms via PatientRecord versus 11.0 ms from raw OMOP, a ~36.8× speedup.

**Discussion:** The projection's significance is as a foundation for other applications: it lowers each one's marginal cost by computing error-prone clinical derivation once and removing it from every consumer. Line-of-therapy inference showed decision-readiness demands embedded clinical reasoning, and that the projection is a living artifact requiring maintenance.

**Conclusion:** A flattened, decision-ready projection over a standards-based longitudinal record is a viable, deployed pattern for turning fragmented data into actionable infrastructure, while remaining OMOP-conformant. Benchmarks measured a ~36.8× eligibility-screening speedup.

## Introduction

Across health systems and biopharma, the hardest problem in applied health AI is rarely the model—it is the data beneath it. Patient records are scattered across providers, electronic health records (EHRs), laboratories, and registries [1,2]. Even where data is captured in standards-conformant form, the mappings are manual, the structure is optimized for storage rather than decision-making, and the distance between holding patient data and acting on it for an individual remains wide and largely hand-built [1,3].

A consequence of this gap is repeated, redundant effort. Each downstream application—a cohort analytics pipeline, a clinical trial matcher, a clinical decision support (CDS) rule engine—must independently reconstruct the patient's clinical state from the underlying transactional record [2,4]: resolving lines of therapy, determining current disease status, normalizing biomarkers, reconciling conflicting source values [5,6]. This re-derivation is expensive, error-prone, and a frequent source of inconsistency between applications that should agree.

This paper presents **PRomop** (PatientRecord on OMOP), an open-source longitudinal patient health record designed to close that gap. PRomop makes two commitments simultaneously: it is standards-based, building on the Observational Medical Outcomes Partnership (OMOP) Common Data Model so that it interoperates with the broader Observational Health Data Sciences and Informatics (OHDSI) ecosystem [7]; and it is decision-ready, exposing a flattened projection of each patient that downstream applications can consume directly without re-deriving clinical state.

The central architectural idea is a deliberate separation between a transactional record and a decision-ready projection. The transactional layer—standard OMOP tables with oncology extensions—preserves the longitudinal source record, including lower-level clinical events and oncology-specific abstractions [8,9]. A computed projection we call *PatientRecord* collapses that history into a single wide row representing what is true now - a decision-oriented view, not a replacement for the source record.Patient-state derivation is performed once, during projection, and materialized; every consuming application reads the same projection rather than reconstructing state.

PatientRecord goes beyond merely retrieving the most recent values from standard OMOP tables; it acts as a home for complex clinical derivation. It computes decision-ready attributes that never surface in the raw transactional data, such as inferring a patient's current line of therapy from unstructured notes and scattered drug exposures [5,6], or calculating composite disease-state markers like MeetsCRAB and MeetsSLiM in multiple myeloma [10].

PRomop is intended to serve as a longitudinal patient record that includes multimodal data, extending beyond diagnoses, medications, labs, procedures, and clinical notes to genomics [11] and imaging (via the Medical Imaging CDM) [12]. This goal also includes supporting wearable data, where missingness, measurement bias, and selection bias require careful metadata and aggregation choices [13], as well as patient-reported outcomes, for which OMOP integration has been explored but remains affected by vocabulary and representation gaps [14].

Our contributions are:

- An architecture that separates a standards-based transactional record from a flattened, decision-ready projection (PatientRecord), enabling multiple application classes to serve from one substrate.
- Oncology extensions for OMOP, including oncology-specific abstractions such as episodes and episode-event links, developed through the OHDSI ecosystem rather than as a proprietary fork [8,9].
- Operational evidence from production deployment across two oncology organizations—the independently governed HealthTree Foundation and the HealthKey-owned CancerBot—totaling ~17,500 patients, including trial matching across 19,500 actively recruiting trials in five cancer types, together with a controlled local benchmark showing that a representative 20-criterion eligibility search reduced from 27–39 joins over raw OMOP to zero joins against the projection and achieved a ~36.8× empirical speedup on a 1000-patient synthetic Synthea-generated breast-cancer cohort.
- Candid deployment lessons, particularly the inadequacy of purely structural approaches for inferring lines of therapy and the ongoing maintenance burden of a demand-coupled projection.

We frame PRomop explicitly as infrastructure: do the hard part—a clean longitudinal record—once, and make everything else a query against it.

## Background and Significance

### *The OMOP Common Data Model and OHDSI*

The OMOP CDM is a widely adopted standard for representing observational health data in a common relational schema, maintained by the OHDSI community [7]. Its strength is interoperability: standardized vocabularies (e.g., SNOMED CT, RxNorm, LOINC) and a shared schema allow analyses written once to run across institutions [7]. OMOP is, however, a model optimized for storage and population-level observational analysis. It is normalized and event-oriented; answering a question about an individual patient's current clinical state typically requires substantial joins, feature extraction, and derivation [15]. PRomop adopts OMOP as its foundation precisely to inherit this interoperability, and adds a projection layer to address decision-readiness for the individual patient. Prior reviews have described OMOP CDM as a basis for harmonized, multicentric data-driven cancer prediction while also noting the need for better support for oncology-specific and predictive use cases [11].

### *Oncology extensions*

Base OMOP represents oncology concepts such as cancer episodes and lines of therapy only partially. The OHDSI Oncology Working Group has developed extensions to the CDM, including the Episode and Episode_Event tables that model disease episodes and treatment regimens [8,9]. PRomop builds on this direction and contributes additional oncology columns, which we intend to submit upstream.

### *Flattened and feature-oriented patient representations*

The tension between normalized clinical models and flattened, analysis- or ML-ready representations is well known; feature stores and denormalized "patient-level" tables are common in practice, and prior work has examined feature extraction and engineering from OMOP-structured and other clinical data for downstream modeling [15]. PRomop's contribution is not the idea of flattening per se, but the discipline of computing a single canonical decision-ready projection once, materializing the derivations, and serving all downstream workloads—analytics, matching, and CDS—from it.

### *Clinical trial matching*

Automated trial matching has seen substantial recent work, much of it using large language models—for example TrialGPT [4], TrialMatchAI [16], PRISM/OncoLLM [17]. Much of this work returns ranked lists of candidate trials. PRomop's companion matcher, EXACT, instead evaluates eligibility criterion-by-criterion against the PatientRecord projection and returns a tri-valued verdict (eligible / potential / ineligible) per trial; a full treatment of EXACT is outside this paper's scope, but it is one of the applications demonstrating the shared-substrate architecture. *(Note: the trial-matching system PRISM [17] is unrelated to PRism, the population-analytics component of the PHRAME suite described in this paper; the similar names are coincidental.)*

## Materials and Methods

PRomop's design rests on a two-layer separation between a transactional record and a decision-ready projection, implemented on open standards and validated through production deployment. Figure 1 summarizes the layers.

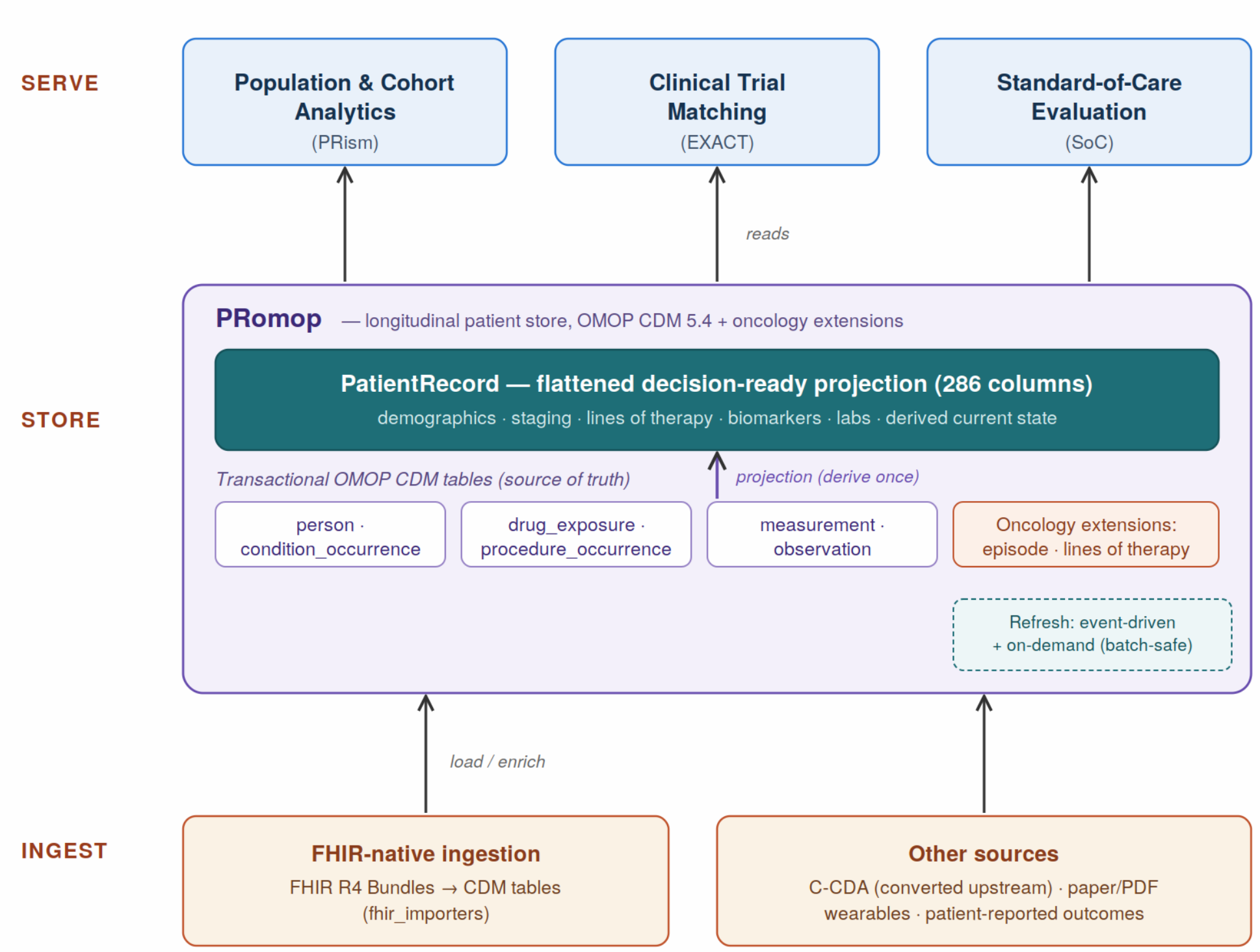

**Figure 1.** *PRomop architecture. An ingest layer (FHIR-native and other sources) loads data into the store layer: transactional OMOP CDM 5.4 tables with oncology extensions serve as the source of truth, from which the PatientRecord projection (304 columns) is derived once. All serve-layer applications—population analytics (PRism), clinical trial matching (EXACT), and standard-of-care evaluation (SoC)—read from the same projection rather than reconstructing patient state.*

### *Storage layer*

We adopt OMOP CDM 5.4 as the foundation, populating its standard clinical tables (person, condition_occurrence, drug_exposure, measurement, observation, procedure_occurrence) and extending it with oncology-specific structures for episodes and lines of therapy that the base model represents only partially [8,9]. Source data is ingested via OHDSI-standard practice: profiling source schemas, authoring explicit field mappings, and mapping vocabularies to LOINC, SNOMED CT, RxNorm, and ICD-O-3 [7]. Keeping PRomop conformant with OMOP avoids forking into a proprietary schema and preserves compatibility with OHDSI tooling [7]. The underlying store and patient portal conform to an oncology Functional Profile of the HL7 Personal Health Record System Functional Model Release 2 (PHR-S FM R2), meeting all 25 essential functions, including tamper-evident, hash-chained audit with delete-restriction, authentication hardening, and versioned terminology management; the profile and its criterion-level traceability matrix are maintained in the PRomop repository.

### *Projection layer: PatientRecord*

The architectural core is *PatientRecord*: a flattened, denormalized projection that collapses each patient's full longitudinal history into a single wide row of 304 columns, of which 299 span six categories: demographics and social context (26 fields), disease and staging (75 fields), laboratory values and biomarkers (119 fields), treatment history and therapy lines (27 fields), derived and computed clinical fields (33 fields, e.g., line of therapy, MeetsCRAB, TNBC status, TP53 disruption), and behavioral and wearable data (19 fields). The fields are not an arbitrary flattening of OMOP; they are the union of patient attributes that the served decision workloads evaluate—the eligibility criteria of the trials matched, the inputs to standard-of-care and guideline rules, and the variables required for cohort analytics—so the categories mirror the axes on which oncology eligibility and guidelines are actually expressed. Within them, priority is given to derived, composite fields that are expensive and error-prone to recompute; a field-by-field mapping of every column, with a per-field classification (direct, selected-latest, aggregated, inferred, or clinically-derived) and the rules governing multiplicity, missingness, conflicts, timestamps, and provenance, is provided in Appendix A (supplementary). Where the transactional tables preserve everything that ever happened, PatientRecord is computed to represent what is true now: demographics, staging, treatment lines, biomarkers, laboratory values, and derived clinical state such as prior therapy and current disease status. PatientRecord is therefore a decision-oriented projection, not a lossless replacement for the longitudinal OMOP record: the transactional layer remains the complete source of truth, and any state the projection omits or collapses is recoverable by re-derivation from it. Because virtually every source relation is one-to-many—repeated measurements, multiple diagnoses, serial drug exposures—the projection collapses each to what clinical guidelines consume: typically the most recent value, a guideline-defined window statistic such as a moving average (e.g., wearable summaries), or a computed composite; counts are used only where a criterion requires one. The projection does not expose event timestamps—the original FHIR-derived times remain in the OMOP tables and are kept current by the same refresh, so temporal relationships are preserved exactly as in the relational store—and data cleaning and conflict resolution occur upstream of OMOP, outside PRomop's scope. Because each field is produced by deterministic derivation code rather than manual abstraction, the OMOP rows underlying any projected value are recoverable, and a per-(row, field) lookup back to those source rows is available on demand.

Derivation logic that would otherwise be re-implemented in every downstream consumer—computing lines of therapy, resolving current status, normalizing biomarkers—is performed once, at projection time, and materialized.

Crucially, this projection is not merely aggregating the most recent values from OMOP tables using simple date filters. It embeds complex clinical reasoning to compute attributes that never surface in the raw transactional schema. For example, deriving a patient's current **line of therapy** requires more than scanning the drug_exposure table; prior work has shown that line-of-therapy identification often requires algorithmic derivation and validation, and PRomop combines regimen-detection approaches with local rules and physician-note review to overcome incomplete data [5,6,18]. Similarly, oncology-specific eligibility flags like **MeetsCRAB** and **MeetsSLiM** for multiple myeloma are composite diagnostic constructs rather than discrete laboratory values [10]. They are composite, multi-domain derivations that require PRomop to concurrently evaluate a patient's recent calcium levels, renal function, hemoglobin counts, and bone-lesion observations across the measurement, condition_occurrence, and observation tables, distilling them into a single definitive Boolean state for downstream consumption.

Projection refresh is *event-driven*: changes to a patient's underlying record trigger regeneration of that patient's projection, keeping the decision-ready view current without each application reconstructing patient state. Refresh can also be invoked *on demand*, and event-driven refresh can be disabled during bulk operations (e.g., large backfills or batch loads), where per-event regeneration would be wasteful; the projection is then rebuilt once at the end of the batch. This combination keeps the projection current under normal operation while remaining efficient under bulk update. This automatic currency is a deliberate contrast with hand-built flat projections, which are stale the moment they are materialized and must be rebuilt by hand; synchronizing the projection with OMOP on every write removes that failure mode.

### *Serving layer*

Because every consuming workload reads the same projection, multiple application classes can operate on one substrate rather than maintaining divergent copies. Examples include four distinct application classes: (i) population and cohort analytics; (ii) per-criterion clinical trial matching; (iii) guideline-based standard-of-care evaluation; and (iv) patient-facing conversational agents. This shared-substrate design is the method's central efficiency claim: the join surface and state-derivation cost that normally scale with the number of applications are collapsed into a single shared step. As a result, additional downstream applications can be added more readily because much of the patient-state retrieval and derivation burden is handled once in the shared projection rather than rebuilt separately from normalized source tables.

### *Ingestion and currency*

Records are enriched over time through FHIR-native ingestion: FHIR R4 resources are routed into the CDM tables [19], while non-FHIR formats (e.g., C-CDA) are converted upstream. As new clinical events arrive, the projection updates and downstream capabilities reflect the richer record automatically, including the resolution of previously incomplete states.

### *Evaluation approach*

We report operational evidence rather than a controlled trial: two organizations—the independently governed HealthTree Foundation and the HealthKey-owned CancerBot—have adopted and operate PRomop in production. Because PRomop is open-source infrastructure that these organizations run under their own data governance, we do not report analyses of their patient-level data—which the authors neither hold nor access—but rather the fact and shape of deployed use: where the projection model held up, where state derivation proved harder than anticipated, and what operational experience revealed about the true cost of decision-readiness. This operational evidence is complemented by a controlled timing benchmark—measuring the 20-criterion eligibility-screening query directly—conducted entirely on a synthetic breast-cancer cohort in a local development environment. The benchmark cohort was generated from Synthea breast-cancer data and imported into the same OMOP-plus-PatientRecord pipeline used elsewhere in PRomop. We then added only the minimum synthetic OMOP rows needed where the generator left gaps that materially affected decision-ready fields—principally histology, HemOnc-coded therapy representation, wearable summaries, and a small number of biomarker or lifestyle observations—so the benchmark exercised the real derivation code rather than a hand-curated flat table. Crucially, this enrichment was applied once to the shared OMOP substrate before either arm was timed, so the PatientRecord and raw-OMOP paths queried identical source data; the enrichment therefore cannot advantage either path or inflate the measured speedup.

## Results

### *Existence proof at scale*

PRomop's primary result is an existence proof at scale: a standards-based longitudinal record with a decision-ready projection, operating in production rather than in prototype. The architecture is deployed by two oncology organizations—the independently governed HealthTree Foundation (~14,000 blood-cancer patients) and CancerBot (~3,500 patients), a HealthKey-owned deployment—totaling roughly 17,500 real patients drawn from fragmented, heterogeneous, real-world sources. This demonstrates that the two-layer design generalizes beyond a single dataset or institution.

### *Decision-readiness in practice*

The PatientRecord projection collapses each patient's longitudinal history into a single 304-column decision-ready row across both deployments. Derivation logic for clinical state—lines of therapy, current disease status, normalized biomarkers—is computed once at projection time rather than re-implemented per application.

The clearest way to quantify the benefit is the cost of eligibility screening, the workload that most directly exercises patient state. Consider a realistic 20-criterion eligibility search over raw OMOP. Such a query spans person, condition_occurrence, measurement, drug_exposure, episode, and procedure_occurrence, with a concept lookup for nearly every clinical criterion. Laboratory criteria are the dominant cost: because measurement stores one row per test per date, each lab criterion needs not only a table join and a concept join but a correlated subquery (effectively GROUP BY with MAX(date)) to recover the most-recent value. The query cost grows roughly as O(n_criteria × n_measurement × log n_concept), and the measurement table dominates cardinality—on the order of 10 –10 rows per 10,000 patients (Table 1).

| Table | Rows per 10k patients |
|---|---|
| measurement | 1M – 10M |
| drug_exposure | 100k – 1M |
| condition_occurrence | 50k – 500k |
| concept (CDM-wide) | 2M+ |
| PatientRecord | 10k (one row per patient) |

**Table 1.** *Approximate table cardinality.*

Against PatientRecord, the same 20-criterion search requires zero joins: every criterion resolves to a predicate (AND col = value) on a single 10,000-row table, because the latest values, concept resolutions, and derived states have already been computed at ingest. Adding a criterion is linear in patient count rather than in criteria count, whereas over raw OMOP each additional criterion adds at minimum one join and typically a subquery. Table 2 summarizes the comparison.

| Approach | Joins | Rows touched |
|---|---|---|
| Raw OMOP, 20 criteria | 27–39 | 5M–15M across tables |
| PatientRecord, 20 criteria | 0 | 10k, one table |

**Table 2.** *Join-count comparison for a 20-criterion eligibility search at 10,000 patients. The gap widens with each added criterion; Table 3 reports a controlled empirical measurement of the underlying mechanism.*

We characterize this as a 30×–200× effective speedup for a 20-criterion eligibility search, with the join count reduced from 27–39 to zero. The range is wide and deliberately so: it reflects dependence on index coverage, PostgreSQL planner choices (nested-loop vs. hash join), and short-circuit evaluation. The benefit grows with criteria count: at 20 criteria, the raw-OMOP query is borderline unexecutable interactively without precisely the kind of pre-computed projection PatientRecord provides.

We further tested the underlying mechanism with a controlled, empirical benchmark run on a local development machine, using a synthetic but OMOP-structurally-representative Synthea-generated breast-cancer patient cohort (n = 1000) built on the same PRomop schema and derivation codebase as the production deployments. To avoid a strawman comparison, the raw-OMOP baseline was executed against tables with standard B-tree indexes on the join and filter keys (person_id, the relevant *_concept_id columns, and event dates), so the live-derivation path reflects a competently indexed schema rather than an unindexed worst case. The cohort was imported as Synthea-generated breast-cancer data; we then selectively enriched only the OMOP rows necessary to make missing decision-ready fields evaluable, rather than hand-authoring PatientRecord rows. We measured wall-clock time to (a) retrieve the 20 trial-eligibility criteria from the materialized PatientRecord row versus (b) fetch the same 20 fields live from the underlying OMOP tables using correlated subqueries—the same query shape described analytically in Tables 1–2—over 5000 timed samples (1000 patients × 5 passes). Both paths populated an identical 15 of 20 eligibility fields on average, confirming the comparison is apples-to-apples. Table 3 reports the results: PatientRecord averaged 0.30 ms (95% CI: 0.29–0.30 ms) versus 11.0 ms (95% CI: 11.0–11.1 ms) for the raw-OMOP path—a ~36.8× mean speedup (95% CI: 36.7–37.0×), consistent with the mechanism of eliminating repeated joins and per-query derivation.

| Metric | PatientRecord | Raw OMOP |
|---|---|---|
| 20-criterion eligibility screening (n = 5000, 1000 patients × 5 passes) | | |
| Mean | 0.30 ms (95% CI: 0.29–0.30) | 11.0 ms (95% CI: 11.0–11.1) |
| Median | 0.29 ms | 10.8 ms |
| p95 | 0.34 ms | 12.4 ms |
| Speedup (mean) | — | 36.8× (95% CI: 36.7–37.0×) |

Table 3. Empirical benchmark of PatientRecord versus raw OMOP for a 20-criterion eligibility screening query, 1000-patient Synthea-generated breast-cancer cohort, run on local hardware (n = 5000 timed samples, 1000 patients × 5 passes). Both paths return an identical 15/20 eligibility fields on average. B-tree indexes on join and filter keys are present in the raw-OMOP path.

### *Code simplification*

The performance gap has a software-engineering counterpart. In the open-source benchmark both paths retrieve the identical 20-criterion row, yet against PatientRecord the fetch is a seven-line single-table read (Figure 2A): because every criterion is an ordinary column, an eligibility filter is a conjunction of predicates reading line-for-line like the trial protocol's inclusion criteria.

The equivalent raw-OMOP fetch runs to roughly 200 lines (Figure 2B): correlated latest-value subquery builders per source table, per-criterion LOINC codes with source-value fallbacks, a keyword filter to find the cancer diagnosis, and normalization functions re-applied after the query returns.

**(A) PatientRecord – the complete 20-criterion fetch**

```
# the complete 20-criterion eligibility fetch
def fetch_trial_row(person_id):
    return (
        PatientRecord.objects
        .filter(person_id=person_id)
        .values(*TRIAL_ELIGIBILITY_FIELDS)  # 20 columns, one table, zero joins
        .get()
    )
```

**(B) Raw OMOP – the equivalent fetch (excerpt)**

```
# excerpt of the equivalent fetch (~200 lines in full)
def _latest_measurement_number_subquery(concept_codes, output_field):
    qs = (
        Measurement.objects
        .filter(person=OuterRef('pk'))
        .filter(
            Q(measurement_concept__concept_code__in=concept_codes)
            | Q(measurement_source_value__in=concept_codes)
        )
        .exclude(value_as_number__isnull=True)
        .order_by('-measurement_date', '-measurement_id')
        .values('value_as_number')[:1]
    )
    return Subquery(qs, output_field=output_field)

def fetch_omop_trial_row(person_id):
    row = (
        Person.objects
        .filter(person_id=person_id)
        .annotate(
            disease=_latest_cancer_condition_subquery(),
            ecog_performance_status=_latest_observation_number_name_subquery(
                ['ecog'], IntegerField()),
            hemoglobin_g_dl=_latest_measurement_number_subquery(
                ['718-7'], DecimalField(max_digits=5, decimal_places=1)),
            serum_creatinine_mg_dl=_latest_measurement_number_subquery(
                ['2160-0', '38483-4'], DecimalField(max_digits=5, decimal_places=2)),
            her2_raw=_latest_measurement_text_subquery(['48676-1']),
            # ... 15 further annotations, one correlated subquery per criterion ...
        )
        .values(...)
        .get()
    )
    return {
        'gender': _normalize_gender(row['gender_raw']),
        'her2_status': _normalize_receptor_status(row['her2_raw']),
        # ... normalization re-applied per field in application code ...
    }
```

**Figure 2.** *The same 20-criterion trial eligibility fetch expressed against the PatientRecord projection (A, complete) and against raw OMOP tables (B, excerpt), taken from the open-source benchmark implementation (benchmark_trial_eligibility.py). The raw-OMOP path additionally requires four further latest-value subquery builders and three normalization functions not shown.*

The reduction matters less for lines saved than for errors avoided. Each raw-OMOP criterion is an independent chance for silent failure—an incomplete concept list, a missing source-value fallback, a mis-ordered latest value, an absent null exclusion—none failing loudly, yet each able to misclassify eligibility.

The projection moves this defect class out of the query path: concept resolution, fallbacks, latest-value selection, and normalization are computed and validated once, then inherited by every consumer rather than re-implemented in each. What remains is short enough to review against the protocol criterion-by-criterion—by a clinician, not only an engineer.

### *One substrate, multiple applications*

The shared-substrate claim held: population analytics, clinical trial matching, and standard-of-care evaluation all run against the same projection without maintaining separate copies of patient state. For population analytics, PRomop supports cohort-level views over the longitudinal record without requiring each analysis to reconstruct current patient status from transactional OMOP tables. For clinical trial matching, PRomop supports criterion-level evaluation against approximately 19,500 actively recruiting trials spanning follicular lymphoma, multiple myeloma, breast cancer, chronic lymphocytic leukemia, and mantle cell lymphoma, and is applied by the deployments across their combined ~17,500-patient population. For standard-of-care evaluation, the same projection provides the current disease, treatment, biomarker, and laboratory context needed to compare patient state against guideline-based recommendations.

Patient-facing conversational agents represent a further potential workload on the same substrate. Empathica, a DCI Network initiative [20], could use PRomop/PatientRecord as a patient-context layer, allowing it to consume both denormalized patient data and precomputed clinical logic—such as line-of-therapy inference and composite disease-state attributes—rather than re-deriving these during conversation.

Across implemented and potential workloads, adding new applications becomes a matter of querying an existing record rather than rebuilding patient state, lowering the marginal cost of each new capability.

### *Operational lessons*

***Inferring lines of therapy.*** This proved a particular challenge. Purely structural normalization was insufficient for real-world oncology data; we supplemented the OHDSI ARTEMIS regimen-detection approach [18] with local rules and physician-note review to derive therapy lines reliably from incomplete and inconsistent sources. This suggests a broader principle: the decision-ready projection is where clinical reasoning, not merely data transformation, must live.

***The projection is a living artifact.*** Decision-readiness is not a stable end state. As new trial-eligibility criteria and CDS rules emerged, additional fields had to be added to the projection, requiring ongoing vigilance to keep the decision-ready view aligned with downstream demand. Teams adopting this pattern should plan for projection maintenance as a continuous obligation. To make that maintenance reproducible, each PatientRecord row carries a derivation_version and a derived_at timestamp recording the logic version that produced it; the version is incremented whenever aggregation logic changes, a backfill command re-derives affected records under the new version, and a derivation changelog documents each change, so any projected value can be reproduced from the OMOP source under a known version of the derivation code.

### *Ecosystem contribution*

A further outcome is standards-facing: the oncology extensions developed for PRomop are being prepared as a proposed upstream contribution to the OHDSI oncology effort, so the work strengthens the shared model rather than fragmenting it.

## Discussion

The central lesson of PRomop is that the hardest part of applied health AI is not the model or even standardization—it is making a standardized record decision-ready, and doing so once rather than repeatedly. Eliminating the 27–39 joins of a raw-OMOP eligibility query down to zero is striking, but its real significance is economic rather than technical: it changes the marginal cost of every new application. When patient state is derived once at projection time, adding a trial matcher, an analytics view, or a CDS rule becomes a query against an existing record rather than a fresh state-reconstruction effort. The architecture's value compounds as applications accumulate.

A natural objection is that PatientRecord is "just" a materialized view, a feature store, or a dbt model over OMOP, and that off-the-shelf tooling for incremental materialization would suffice. Mechanically, that is accurate: PatientRecord is a materialized denormalization with event-driven refresh, and it could be implemented with any of those technologies. The objection misses where the work is. Related alternatives exist but address different problems. FHIR bulk export and flat-file representations (e.g., ndjson FHIR, Parquet-serialized resources) are optimized for data transfer and interoperability, not for direct queryability or embedded clinical derivation; they still require consumers to reconstruct patient state from normalized event resources. These approaches do not materialize clinical reasoning such as line-of-therapy inference or composite disease-state flags. A materialized view or dbt model automates the mechanical parts—joining, aggregating to the latest value, refreshing on change—while leaving the clinical derivation to each consumer. The fields that matter most downstream are not simple latest-value rollups but validated clinical reasoning: line-of-therapy inference from scattered drug exposures and unstructured notes, composite disease-state constructs such as MeetsCRAB and MeetsSLiM, and biomarker normalization across heterogeneous sources. PatientRecord's contribution is to make that derivation a first-class, once-computed, validated artifact rather than an error-prone task that each developer or data scientist re-implements—differently, and with divergent results—inside every analytics pipeline, matcher, and CDS rule. The materialization technology is interchangeable; the discipline of computing clinical decision-readiness once, and removing that error-prone step from every downstream team, is the point.

The lines-of-therapy experience complicates any claim that decision-readiness is purely a structural problem. The fields hardest to derive are precisely those most valuable downstream, and they resist tidy extract-transform-load; they require embedded clinical reasoning, including reading unstructured notes. The stakes are not merely performance: without a correctly derived line of therapy, eligibility for the majority of oncology trials cannot be determined at all, because most eligibility criteria are expressed in terms of prior treatment requirements—‘received at least one prior line' or 'no more than three prior regimens'—making the derived field a prerequisite for any downstream trial-matching evaluation. A second lesson tempers the architecture's appeal: the same property that makes PatientRecord powerful—pre-computing what consumers need—makes it a living artifact coupled to evolving clinical demand, with a continuous maintenance cost.

These results carry limitations worth stating plainly. Our evidence for production operation is operational rather than benchmarked: we demonstrate that the architecture works at scale across two organizations. The controlled timing benchmark we report (Table 3) was run on local hardware against a synthetic Synthea-generated cohort rather than the two named production deployments directly, and it measured one representative workload: a 20-criterion eligibility-screening query. The synthetic cohort was selectively enriched where needed to fill generator gaps in a small number of decision-ready fields; this improved completeness while preserving the benchmark's dependence on the same OMOP derivation path used in PRomop itself. Both deployments are in oncology; while the OMOP foundation is disease-agnostic, the oncology extensions and derived fields are not yet validated outside this domain. The 30×–200× analytical estimate and the ~36.8× empirical measurement should be read as illustrative of the pattern's effect for this query shape and cohort rather than as a universal benchmark across criteria mixes or deployments. The measured ratio scales with the size of the OMOP measurement table: a smaller 100-patient pilot (~17,000 measurement rows) yielded ~24×, whereas the 1000-patient cohort reported here (~217,000 rows) yields ~36.8×, moving into the analytical band as the table approaches—though remains below—the 1M–10M rows assumed at production scale. This scaling is consistent with the mechanism, since the raw-OMOP path's cost grows with measurement-table cardinality while the constant-time PatientRecord lookup does not. Finally, because both deployments share architectural lineage, the independence of replication is partial.

Situated against the field, PRomop's contribution is deliberately not novelty in the data model—it builds on OMOP precisely to avoid fragmenting the standards ecosystem and returns its oncology extensions upstream. The novelty is the projection layer and the discipline of computing decision-readiness once. We believe this pattern generalizes beyond oncology and beyond our deployments, and that operational evidence of the kind presented here is the appropriate contribution: not a proof that the method is optimal, but a demonstration that the pattern is viable, deployed, and serving real patients.

## Conclusion

PRomop demonstrates that a flattened, decision-ready projection over a standards-based longitudinal record is a viable, deployed pattern for turning fragmented patient data into infrastructure that AI applications can act on. By computing patient state once and serving analytics, trial matching, and standard-of-care evaluation from a single shared record, PRomop collapses the redundant derivation that burdens conventional pipelines—reducing a 20-criterion eligibility search from 27–39 joins over raw OMOP to zero against the projection, an estimated 30×–200× speedup—while remaining conformant with OMOP and contributing its oncology extensions back to OHDSI. In a controlled benchmark on a 1000-patient Synthea-generated breast-cancer cohort run on local hardware, PatientRecord averaged 0.30 ms versus 11.0 ms for 20-criterion eligibility screening (~36.8×, 95% CI 36.7–37.0×). Deployed across ~17,500 real oncology patients, it offers a concrete, open-source pattern for moving from proof to practice. Future work includes validation beyond oncology, repeated benchmarking across broader eligibility criteria mixes and larger cohorts, and completion of the upstream OHDSI oncology contribution.

## Ethics and Human Subjects

This work describes open-source software infrastructure. The authors developed and released PRomop; the deploying organizations (the HealthTree Foundation and CancerBot) operate it independently, under their own data governance, and use the software as they see fit. The authors did not access, analyze, or measure identifiable patient data at those organizations. The production figures reported here—patient counts, trial counts, and deployed capabilities—reflect the organizations' use of the software rather than analyses performed by the authors. The controlled timing benchmark (Table 3) was conducted entirely on a synthetic Synthea-generated breast-cancer cohort, with limited additional synthetic enrichment only where source gaps prevented evaluation of decision-ready fields, and contained no real patient data. Accordingly, the work reported here did not constitute human subjects research and did not require institutional review board approval.

## Competing Interests

A.B. is co-founder and Chief Technology Officer of HealthKey, Inc., which develops PRomop and the PHRAME suite and owns CancerBot. The HealthTree Foundation holds an equity interest in HealthKey, Inc. L.F.-A. declares no competing interests. The authors otherwise declare no competing financial or non-financial interests.

## Funding

This work was supported by service and infrastructure fees paid to HealthKey, Inc. by the HealthTree Foundation, and by HealthKey, Inc. No external grant funding was received.

## Data Availability

The PRomop source code and the PHRAME components are openly available at https://github.com/healthkey-ai, and a live analytics demonstration on synthetic data is available at https://prism.healthkey.ai. The synthetic Synthea-generated breast-cancer cohort underlying Table 3 is archived at Zenodo (DOI: https://doi.org/10.5281/zenodo.21430170). The patient-level data underlying the production deployments cannot be shared: it is protected health information held by the deploying organizations under their own governance, and the authors neither hold nor control it.

## Reproducibility

The 1000-patient synthetic breast-cancer cohort underlying Table 3, including the targeted synthetic enrichment applied where the Synthea generator left decision-ready fields empty, is archived as synthea_bc_1000.json at Zenodo (https://doi.org/10.5281/zenodo.21430170); the smaller 100-patient pilot cohort referenced in the Discussion remains available as

synthea_bc.json (https://zenodo.org/records/21312915). The cohort can be imported into a PRomop deployment via the import_org_patients management command, after which the benchmark_trial_eligibility command reproduces Table 3 for the 20-criterion eligibility-screening query directly comparable to the query shape in Tables 1–2. Complete setup instructions, command options, and the OMOP-to-PatientRecord field mapping exercised by the live-derivation path are documented in the PRomop repository (docs/reproducing-benchmark-results.md). Because query latency is hardware- and cache-dependent, the published millisecond figures reflect the specific environment in which they were measured; the relative speedup between the PatientRecord and live-OMOP paths is the result intended to be reproducible.

## Author Contributions

A.B.: Conceptualization, Software, Methodology, Investigation, Writing—original draft, Writing—review & editing. L.F.-A.: Methodology, Validation, Investigation, Writing—review & editing. S.L.: Writing-review & editing. All authors read and approved the final manuscript.

## Acknowledgments

We thank the HealthTree Foundation for funding and feedback, and advisors Drs. Steven Labkoff and Yuri Quintana of Harvard Medical School for guidance and insight as we developed this work.